\documentstyle[12pt]{article}
\title{
\vspace{-1.4cm} \hspace{13.7cm} \large PHE 90-17 
\\ \vspace{.5cm} \LARGE
QED corrections with partial angular integration
\\        to fermion pair production in \ee - annihilation}
\author{\vspace{0.cm}\\
         D. Bardin, M. Bilenky, A. Sazonov, Yu. Sedykh\\
          Joint Institute for Nuclear Research, Dubna,
  \\ Head Post Office P.O. Box 79, SU - 101000 Moscow, USSR 
\vspace{-0.4cm}\\
 \\and \and T. Riemann, M. Sachwitz
    \\ Institut f\"ur Hochenergiephysik,
    \\   Platanenallee 6, DDR-1615 Zeuthen/Brandenburg\thanks{Address to be
expected after 3 October 1990: O-1615 Zeuthen/Brandenburg(FRG)}}
 
\date{October 2, 1990}
 
\newcommand{\ee}{e$^{+}$e$^{-}$}
\newcommand{\afb}{$A_{FB}\:$}
\newcommand{\st}{$\sigma_{T}\:$}

\begin{document}
\pagestyle{empty,textwidth,textheight}
\maketitle
%
\begin{abstract}
    Analytic formulae are derived for the complete photon energy 
    spectrum due to
    QED corrections to fermion pair production in case of a limited 
    angular acceptance for the final state fermions. After a numerical
    integration over 
    the energy of non-observed photons, this corresponds to 
    typical experimental conditions at LEP/SLC. 
\end{abstract}
\vspace{1.cm}
%
%
\section{Introduction}
One of the main tasks of the \ee \ storage rings LEP and SLC is a
verification of the standard electroweak theory with unpreceded precision.
An important reaction in this context is fermion pair production:
\begin{equation}
e^{+} + e^{-}
 \longrightarrow ( \gamma ,Z) \longrightarrow f^{+}
+ f^{-} + n\gamma,\hspace{1cm} f \ne e.
\vspace{.5cm}
\label{eq:cross}
\end{equation}
Within the standard theory, the corresponding cross sections can 
be predicted with a precision well below 1 \% either with (semi-)
analytic formulae 
$^{\cite{yr:ber,yr:boe}}$
or Monte-Carlo simulation 
$^{\cite{yr:kle}}$. For the analysis
of experimental data, the use of (semi-)analytic formulae is preferred
due to their fast performance and well understood accuracy. A 
disadvantage is the small flexibility concerning the choice of 
experimental cuts. 
 
      In a series of papers, we derived analytic formulae taking into account 
QED corrections to the total cross section and the integrated 
forward-backward asymmetry 
as well as to the differential cross-section (see 
$^{\cite{own.phe:allqed}}$ and references quoted therein).
The demand of experimentalists for fast and elegant algorithms 
motivated us to derive in addition 
analytic formulae for the photon energy spectrum in presence of
an angular acceptance cut for the final state fermions. Such formulae
allow a description of the quite realistic situation of fermion pair
production including photon bremsstrahlung (with soft photon exponentiation)
with cuts on the photon energy and on the fermion scattering angle using
only one numeric integration. These new formulae are the subject of the present
article. 
 
     We denote with $\sigma_{T}(-c_{1},c_{2})$ and $A_{FB}(-c_{1},c_{2})$
the total cross section and 
forward-\-back\-ward asymmetry, both integrated over the angular range 
$cos\theta  \in  (-c_{1},c_{2})$:
\begin{equation}
\sigma_{T}(-c_{1},c_{2}) = \sigma_{T}(c_{2}) - \sigma_{T}(-c_{1}),
\label{eq.1.1}
\end{equation}
\begin{equation}
A_{FB}(-c_{1},c_{2}) = \sigma_{T}(-c_{1},c_{2})^{-1}
[\sigma_{FB}(c_{2}) + \sigma_{FB}(-c_{1})].
\label{eq.1.2}
\end{equation}
  The building blocks $\sigma_{A}(c),A=T,FB$ have to be
 determined by an explicit
integration over the angular distribution:
\begin{equation}
   \sigma_{A}(c) = \sum_{m,n=0,1} \;\sum_{a=e,i,f} Re [d_{A} \:
    \sigma_{A}^{a,0}(s,s;m,n) \:R_{A}^{a}(c;m,n)],
\label{eq.1.3}
\end{equation}
\begin{equation}
R_{A}^{a}(c;m,n) = d_{A}^{-1} \int_{0}^{\Delta} dv \:R_{A}^{a}(c,v;m,n)
    \frac{\sigma_{A}^{a,0}(s,s';m,n)}{\sigma_{A}^{a,0}(s ,s ;m,n)},
        \label{eq.1.4}
\end{equation}
\begin{equation}
d_{T} = \frac{4}{3}, \hspace{1cm} d_{FB} = 1, \hspace{1cm} s'=(1-v)s.
\end{equation}
Here, $\sigma_{A}^{a,0}(s,s';m,n)$ are normalisation
factors containing the Born resonance function and all the couplings.
Indices $m,n$ are for photon $(m=0)$ and $Z$ boson $(m=1)$ exchange.
The remaining (numerical) integration in (~\ref{eq.1.4}) is to be 
performed over
the photon energy with cut-off $\Delta \in (0,1)$. The problem to be 
solved is the determination of the angular dependent functions,
\begin{equation}
R_{A}^{a}(c,v;m,n) = \int_{0}^{c} dcos\theta \:r_{A}^{a}(cos\theta,v;m,n).
\label{eq.1.5}
\end{equation}
\sloppy
The general notations have been taken over from 
$^{\cite{own.phe:allqed}}$,
where also the functions $r_{A}^{a}(cos\theta,v;m,n)$ may be read
off.
\fussy
 
      The following chapters are devoted to an explicit representation
of rather compact formulae for the QED corrections (~\ref{eq.1.5}) 
originating from initial $(a=e)$ 
and final $(a=f)$ state radiation and their interference $(a=i)$. 
\vfill\eject
\nopagebreak
\section{Initial State Radiation} 
   The major part of radiative corrections originates from the initial state.
Therefore, an inclusion of higher order contributions
is necessary. An
adequate accuracy for LEP/SLC physics may be achieved if one combines the
exact to order $O(\alpha)$ cross sections with a soft photon exponentiation 
procedure
$^{\cite{yr:ber,yr:boe}}$. 
We split the QED corrections into two parts containing soft and hard 
bremsstrahlung, correspondingly. For initial state radiation, they both are 
independent of the kind of exchanged vector boson: 
 \begin{equation}
   R_{A}^{e}(c,v;m,n) = 
   d_{A}(c) [ 1 + \bar{S}(\beta_{e})]
  \beta_{e} v^{\beta_{e}-1} + \bar{H}_{A}^{e}(c,v),
\end{equation}
\begin{equation}
d_{T}(c) =  d_{T}^{-1}(c + c^{3}/3),  \hspace{1cm}  d_{FB}(c) = c^{2},
\label{eq:d33}
\end{equation}
\begin{equation}
 \bar{S}(\beta_{e}) = \frac{3}{4} \beta_{e} +
\frac{\alpha}{\pi} Q_{e}^{2} \left(
 \frac{\pi^{2}}{3}
  - \frac{1}{2} \right),
\end{equation}
\begin{equation}
 \bar{H}_{A}^{e}(c,v) = \frac{\alpha}{\pi} Q_{e}^{2} d_{A}^{-1}
h_{A}^{e}(c,v) / v
- \frac{\beta_{e}}{v} d_{A}(c) ,
\hspace{1cm} A = T,FB,
\label{eq.2.4}
\end{equation}
\begin{equation}
   \beta_{e} = \frac{2\alpha}{\pi} Q_{e}^{2} (L_{e} - 1),
   \hspace{1cm} L_{e} = \ln \frac{s}{m_{e}^{2}}, \hspace{1cm}
   Q_{e}=-1.
\end{equation}
   The second term in (~\ref{eq.2.4}) is the remnant of soft photon 
exponentiation and makes the hard photon correction infra-red finite.
The hard radiator functions are different for \st and \afb.
They depend on the reduced photon energy $v$ and the range of observed 
scattering angles $c$:
\begin{eqnarray}
 h_{T}^{e}(c,v) = \frac{4}{3}r_{2}
  \left[ \frac{\ln\gamma_{+}}{\gamma^{3}_{+}}(c^{3}_{+}-z^{3}c^{3}_{-})
-        \frac{\ln\gamma_{-}}{\gamma^{3}_{-}}(c^{3}_{-}-z^{3}c^{3}_{+})\right]
    \nonumber \\
+ \left.\frac{2c}{\gamma^{3}_{-}\gamma^{3}_{+}}
\right\{z r_{2}[(L_{e}-1)-\ln z] \left[\frac{2}{3}z^{2}(1-c_{+}c_{-})
+  v^{2}c_{+}c_{-}(z+r_{2}c_{+}c_{-})\right]
 \nonumber \\
 +  2 v^{2}z(c_{+}c_{-})^{2}\left(r_{2}z-r_{4}
   + \frac{4}{3} \frac{z^{2}}{c_{+}c_{-}} - \frac{22}{3}z^{2}\right)
+   2 v^{4}(c_{+}c_{-})^{3} \left(\left.
    \frac{5}{3} r_{2}z - 2 z^{2} - \frac{1}{3}r_{4}\right)\right\} ,
\label{eq.2.5}
\end{eqnarray}
\begin{eqnarray}
    h_{FB}^{e}(c,v) = 8z\frac{r_{2}}{r_{1}^{2}}\ln\frac{r_{1}}{2}
  +  \left.\frac{2z}{\gamma^{2}_{-}\gamma^{2}_{+}}\right\{
  [(L_{e}-1)-\ln z][2r_{2}\frac{z^{2}}{r_{1}^{2}}
 \nonumber \\
    - r_{2}^{2} c_{+}c_{-} + 4r_{2}c_{+}c_{-}\frac{v^{2}}{r_{1}^{2}}
   ( z + r_{2}c_{+}c_{-})]
 \nonumber \\
  + v^{2}c_{+}c_{-}[4z - 4c_{+}c_{-}r_{1}^{2} + v^{2}]
  \left\} 
  - 4zr_{2}c_{+}c_{-}(\frac{\ln \gamma_{-}}{\gamma_{-}^{2}}\right.
 +\frac{\ln \gamma_{+}}{\gamma_{+}^{2}}) .
\label{eq.2.6}
\end{eqnarray}
The following abbreviations are used:
\begin{equation}
 r_{n} = 1 + z^{n},  \hspace{1cm} v = 1 - z,
\end{equation}
\begin{equation}
    c_{\pm} = \frac{1}{2}(1 \pm c) ,\hspace{1cm} \gamma_{+} = c_{+} + zc_{-} ,
    \hspace{1cm} \gamma_{-} = c_{-} + zc_{+} .
\label{eq:c311}
\end{equation}
   For an unrestricted angular acceptance, $c\rightarrow1$, the above 
formulae
agree with those derived earlier for the uncut convolution kernels \st 
$^{\cite{bom.np,np:gps}}$ 
and \afb 
$^{\cite{own.pl.afb}}$.
%
%
\section{Initial-Final State Interference Radiation}
   Under usually realised experimental conditions, the $O(\alpha)$ 
initial-final 
interference bremsstrahlung is nearly negligible in the $Z$ resonance region 
$^{\cite{cp:bkj,own.e2:324,pl:jaw}}$.
This is not true if a tight cut on the photon energy is applied or in case 
of scattering angles near $\cos \theta \approx 1$. After an angular 
integration, the second type of singularity is removed completely. 
Away from the peak, the
interference contributions become comparable to the other non-leading 
radiative corrections.
 
    It is sufficient to present the diagonal interference corrections. 
The non-diagonal terms obey the following simple relation:
\begin{equation}
 R_{A}^{i}(c,v;m,n) = \frac{1}{2}[R_{A}^{i}(c,v;m,m) + R_{A}^{i}(c,v;n,n)^{*}],
\hspace{1cm} A = T,FB.
\label{eq:r41}
\end{equation}
The initial-final state interference contributions are composed of soft and 
hard bremsstrahlung parts $S_{A}^{i}(c,\epsilon), h_{A}^{i}(c,v)$ and of 
$\gamma\gamma$ and $\gamma Z$ box diagrams $B_{A}(c;m,n)$:
\begin{equation}
 R_{A}^{i}(c,v;m,n) = \frac{\alpha}{\pi}Q_{e}Q_{f}\{\delta(v)[S_{A}^{i}(c,
 \epsilon) + B_{A}(c;m,n)] + \theta(v-\epsilon)h_{A}^{i}(c,v)
\sigma_{\bar{A}}^{0}(s,s';m,n)\},
\label{eq:s80}
\end{equation}
where $\bar{A} = T,FB$ if $A = FB,T$. The box contributions are 
the only ones with an explicit dependence on the kind of exchanged vector 
boson; they are written in (anti-)symmetrised form:
\begin{equation}
 B_{T,FB}(c;m,n) = b_{T,FB}(c;m,n) 
                             \pm b_{T,FB}(-c;m,n).
\end{equation}
 
    The corrections $R_{T}^{i}(c,v;m,n)$ to the total cross 
section $\sigma_{T}(c)$ contain the following terms:
\begin{eqnarray}
  S_{T}^{i}(c,\epsilon) = -4\ln \epsilon[(c^{2}-1)\ln\frac{c_{+}}{c_{-}} + 2c]
 + 2(c^{2}-1)[Li_{2}(c_{+}) - Li_{2}(c_{-})]
 \nonumber \\
 -(c^{2}-1)\ln(c_{+}c_{-})\ln\frac{c_{+}}{c_{-}}
 - 4c\ln(c_{+}c_{-}) - 4 \ln\frac{c_{+}}{c_{-}} + 8c,
\end{eqnarray}
\begin{eqnarray}
b_{T}(c;0,0)  =& \frac{1}{2}(c^{2}-1)\ln^{2}c_{+} + 2c\ln c_{+}
      - (c^{2}-3)\ln c_{+} - 3c
 - i[\pi (c^{2}-1) \ln c_{+}],
\end{eqnarray}
\begin{eqnarray}
 b_{T}(c;n,n) = -2\ln(1-\frac{1}{R})\{2c R (1-R)\ln c_{+}
 \nonumber \\
 - [-2R^{2} + R(c^{2}+1)+c^{2}-1]\ln c_{+} - c R (R+1)\}
 \nonumber \\
 + 4  R (R-1)c Li_{2}(1-\frac{1}{R}) - 2c\ln R + (c^{2}-1)\ln^{2}c_{+}
 \nonumber \\
 + 4c\ln c_{+} + 2[R(c^{2}-1)-c^{2}+3]\ln c_{+} 
- 4c R(R-1) Li_{2}(1-\frac{c_{+}}{R})
 \nonumber \\
- 2 [2R^{2}-R(c^{2}+1)] Li_{2}(1-\frac{c_{+}}{R}) + 2Rc - 6c
,\hspace{1cm}     n \neq 0,
\end{eqnarray}
 
\begin{equation}
   R = \frac{1}{s}[M_{Z}^{2}-iM_{Z}\Gamma_{Z}(s)],
\end{equation}
\begin{eqnarray}
h_{T}^{i}(c,v) =
 4c_{+}c_{-}\left[\frac{r_{1} r_{2}}{v}\ln\frac{c_{+}}{c_{-}}
  + \right(\frac{c^{2}}{c_{+}c_{-}}z - v^{2}\left)
\ln\frac{\gamma_{+}}{\gamma_{-}}\right]
+ 4cz\left(-\frac{r_{1}}{v} + \ln z\right).
\end{eqnarray}
 
    The interference corrections  $R_{FB}^{i}(c,v;m,n)$ to the numerator of the
forward-\-back\-ward asymmetry, $\sigma_{FB}(c)$ , contain the following terms:
\begin{eqnarray}
  \frac{3}{4}S_{FB}(c,\epsilon) = \ln\epsilon[-8\ln 2 - 4 \ln(c_{+}c_{-})
  - 3 (\frac{c^{3}}{3} + c)\ln\frac{c_{+}}{c_{-}} - c^{2}]
 \nonumber \\
  + \frac{3}{2}(\frac{c^{3}}{3}+c)[Li_{2}(c_{+}) - Li_{2}(c_{-})]
  + 2[Li_{2}(c_{+}) + Li_{2}(c_{-})]
  - \frac{3}{4}(\frac{c^{3}}{3}+c)\ln(c_{+}c_{-})\ln\frac{c_{+}}{c_{-}}
 \nonumber \\
  - (\ln^{2}c_{+} + \ln^{2}c_{-})
  + 4\ln^{2}2 + \ln 2 - \frac{1}{2}\ln(c_{+}c_{-})(c^{2}-1) + \frac{c^{2}}{2}
  - 2 Li_{2}(1),
\end{eqnarray}
\begin{eqnarray}
b_{FB}(c;0,0) 
 = -\frac{1}{2}(c^{2}-1)\ln^{2}c_{+} + (c^{2}-3)\ln c_{+}
      - 2c\ln c_{+} - \frac{1}{2}(\ln^{2}2 + 6\ln 2 + c^{2})
 \nonumber \\
   - i \pi [\frac{5}{3} \ln 2 + (c^{2} 
+ \frac{5}{3})\ln c_{+} + 2 (\frac{c^{3}}{3} + c)\ln c_{+} - \frac{2}{3}c^{2}],
\end{eqnarray}
 
\begin{eqnarray}
 b_{FB}(c;n,n) = (\frac{c^{3}}{3} + c)\ln^{2}c_{+} 
  + \frac{4}{3}\ln^{2}c_{+} 
 \nonumber \\
+ \ln(1-\left.\frac{1}{R})
\right\{(4R^{2} - 2R + \frac{10}{3}) \ln 2
+ [4R^{2}-2R(c^{2}+1)+2c^{2}+\frac{10}{3}]\ln c_{+}
 \nonumber \\
  + 4[cR(R-1)+(\frac{c^{3}}{3}+c)]\ln c_{+} + c^{2}(-R^{2}+3R
-\left.\frac{4}{3})\right\}
 \nonumber \\
 + c^{2}\left(\frac{4}{3}R-\frac{5}{3}\right)\ln R  
+   \left(\frac{16}{3}R^{3} - 4R^{2} + 2R -\frac{2}{3}\right)
 Li_{2}\left(1-\frac{1}{2R}\right)
 \nonumber \\
 + 2c^{2}(R-1) Li_{2}\left(1-\frac{1}{R}\right)  
 - \frac{4}{3}\ln^{2}2 + (\frac{8}{3}R^{2}+\frac{8}{3}R - 6)\ln 2
 \nonumber \\
+\left[\frac{8}{3}R^{2} + R(-\frac{4}{3}c^{2}+\frac{8}{3})
+ 2(c^{2}-3)\right] \ln c_{+} 
 + \left(\frac{8}{3}R^{2}+\frac{4}{3}R-4\right)c\ln c_{+} 
 \nonumber \\
 + 2\left[-\frac{8}{3}R^{3} + 2R^{2}-R(c^{2}+1)+c^{2}
 +\frac{1}{3}\right] Li_{2}\left(1-\frac{c_{+}}{R}\right)
 \nonumber \\
 + 2\left[2c(R^{2}-R) + (\right.\frac{c^{3}}{3}\left.+c)\right]
Li_{2}\left(1-\frac{c_{+}}{R}\right)
 + (\frac{2}{3}R-1)c^{2}
,\hspace{1cm}
     n \neq 0,
\end{eqnarray}
\sloppy
\begin{eqnarray}
h_{FB}^{i}(c,v) =
 \frac{2c^{2}z}{\gamma_{+}\gamma_{-}}
\left(\frac{1}{3} v^{2} - \frac{2}{3 v} + \frac{2 z}{r_{1}}
+\frac{1}{3}c^{2} v r_{1} \right)
\nonumber  \\
 - 2c^{2}z v\ln z -4\frac{r_{1}}{v}c_{+}c_{-}r_{2}\ln (c_{+}c_{-})
\nonumber  \\
 -\frac{16}{3}\frac{r_{3}}{v}(c_{-}^{3}\ln c_{-} + c_{+}^{3}\ln c_{+})
 + \left.\frac{2}{3}\right(2z-5r_{2}\left)\frac{r_{1}}{v}\right.\ln r_{1}
\nonumber  \\
 - 4\ln\gamma_{-} \left[ 2cc_{-}^{2} - \gamma_{-}(v + \frac{4}{3}
 \frac{\gamma_{-}^{2}}{v} - 4c_{-}^{2} + \gamma_{-}) \right]
\nonumber  \\
 + 4\ln\gamma_{+} \left[ 2cc_{+}^{2} + \gamma_{+}(v + \frac{4}{3}
 \frac{\gamma_{+}^{2}}{v} - 4c_{+}^{2} + \gamma_{+}) \right] . 
\end{eqnarray}
\fussy
 
   Again, for an unrestricted angular acceptance the above formulae yield 
those derived earlier $^{\cite{own.pl.afb}}$.
\section{Final State Radiation} 
The final state radiator functions for the angular distribution, 
$r_{A}^{f}(cos\theta,v;m,n)$, as used in (~\ref{eq.1.5}) for the definition
of $R_{A}^{f}(c,v;m,n)$ are described in detail in chapter 5.1 of
$^{\cite{own.phe:allqed}}$. Their angular dependence to order $O(\alpha)$
is the same as that of the Born cross section. Consequently, an
integration is trivial. For applications which deserve highest precision, 
a common exponentiation of initial and final state radiation corrections 
is recommended (see also $^{\cite{pl:nit,br:gre}}$). An inspection
of the formulae in chapter 5.3 of $^{\cite{own.phe:allqed}}$ shows that
the only nontrivial angular dependence is due to the initial state hard 
radiator function. The result of their integration may be found in section 2 of
the present article.
Further, there exists a minor angular variation in the non-leading 
part of 
the final state factor which is proportional to $[3-4/(1+cos^{2}\theta)]$. 
We have shown numerically that this term can be neglected completely for all 
applications. 
 
   As a result, a description of common soft photon exponentiation of
initial and final state bremsstrahlung is obtained by a direct combination
of the treatment in $^{\cite{own.phe:allqed}}$ with the integrated 
expressions for the initial state hard photon corrections (~\ref{eq.2.5}),
(~\ref{eq.2.6}).
\vspace{1.cm}
\section{Discussion}
 
   The formulae of this article may be used both in a model-independent way 
with input parameters chosen 
to be e.g. $M_{Z},\Gamma_{Z},\Gamma_{e},\Gamma_{f}$, or within the standard
theory using e.g. $\alpha,G_{\mu},M_{Z}$ $^{\cite{cp:bar,zp.bhr}}$.
The corresponding Fortran codes \footnote{Besides the calculational chain 
based on the formulae of this article, these codes contain also a branch
with different choice of cuts $^{\cite{bs.e2}}$.}
 ZBIZON and ZFITTER are described in 
$^{\cite{own.note:zfi}}$  and have been applied recently to data obtained 
by LEP experiments (e.g.
$^{\cite{l3.pl:??,aleph.pl:??,delphi.pl:??,opal.pl:??}}$ ).
In a unique way, one can perform a semi-analytical calculation of
either observables which are integrated over a wide
angular range or of pseudo-differential distributions with bins filled using
differences of angular slices. As typical examples, we show in Figs. 1 and 2
some predictions for such distributions of \st and \afb. 
 
    To summarise, we think that the analytic calculation of QED corrections 
with partial angular integration could prove to be a powerful tool for the
phenomenological analysis of fermion pair production.
\pagebreak

\vfill\eject
\listoffigures 
Fig. 1. 
\\
The total cross section $\sigma_{T}(c_{1},c_{2})/\Delta c$, 
$\Delta c=c_{2}-c_{1}$, as function of the scattering angle and of
the photon energy cut-off $\Delta=2E_{\gamma}/ \sqrt{s}$; 
M$_{H}$ = 100 GeV,
m$_{t}$ = 100 GeV.
\\
\\
Fig. 2. 
\\
Cumulative forward-backward asymmetry $A_{FB}(-c,c)$ as a function of
c and of the photon energy cut-off; parameters as in fig. 1.
 
\vfill\eject
\end{document}